# Societal, Economic, Ethical and Legal Challenges of the Digital Revolution: From Big Data to Deep Learning, Artificial Intelligence, and Manipulative Technologies[1]

by Dirk Helbing (ETH Zurich)

*In the wake of the on-going digital revolution, we will see a dramatic transformation of our economy and most of our societal institutions. While the benefits of this transformation can be massive, there are also tremendous risks to our society. After the automation of many production processes and the creation of self-driving vehicles, the automation of society is next. This is moving us to a tipping point and to a crossroads: we must decide between a society in which the actions are determined in a top-down way and then implemented by coercion or manipulative technologies (such as personalized ads and nudging) or a society, in which decisions are taken in a free and participatory way and mutually coordinated. Modern information and communication systems (ICT) enable both, but the latter has economic and strategic benefits. The fundaments of human dignity, autonomous decision-making, and democracies are shaking, but I believe that they need to be vigorously defended, as they are not only core principles of livable societies, but also the basis of greater efficiency and success.*

"Those who surrender freedom for security[2] will not have, nor do they deserve, either one."

<div style="text-align:right">Benjamin Franklin</div>

## Overview of Some New Digital Technology Trends

### Big Data

In a globalized world, companies and countries are exposed to a harsh competition. This produces a considerable pressure to create more efficient systems - a tendency which is re-inforced by high debt levels.

Big Data seems to be a suitable answer to this. Mining Big Data offers the potential to create new ways to optimize processes, identify interdependencies and make informed decisions. There's no doubt that Big Data creates new business opportunities, not just because of its application in marketing, but also because information itself is becoming monetized.

---

[1] This document includes and reproduces some paragraphs of the following documents: "Big Data - Zauberstab und Rohstoff des 21. Jahrhunderts'' published in Die Volkswirtschaft - Das Magazin fürWirtschaftspolitik (5/2014), see http://www.dievolkswirtschaft.ch/files/editions/201405/pdf/04_Helbing_DE.pdf; for an English translation see chapter 7 of D. Helbing (2015) Thinking Ahead - Essays on Big Data, Digital Revolution, and Participatory Market Society (Springer, Berlin).

[2] I would add "efficiency" or "performance" here as well.

Technology gurus preach that Big Data is becoming the oil of the 21st century, a new commodity that can be tapped for profit. As the virtual currency BitCoin temporarily became more valuable than gold, it can be even literally said that data can be mined into money in a way which would previously have been considered a fairy tale. Although many Big Data sets are proprietary, the consultancy company McKinsey recently estimated that the additional value of Open Data alone amounts to be $3-5 trillion per year.[3] If the worth of this publicly available information were evenly divided among the world's population, every person on Earth would receive an additional $700 per year. We now see Open Government initiatives all over the world, aiming to improve services to citizens while having to cut costs. Even the G8 is pushing for Open Data as this is crucial to mobilize the full societal and economic capacity.[4]

The potential of Big Data spans every area of social activity, from the processing of human language and the management of financial assets, to the harnessing of information enabling large cities to manage the balance between energy consumption and production. Furthermore, Big Data holds the promise to help protect our environment, to detect and reduce risks, and to discover opportunities that would otherwise have been missed. In the area of medicine, Big Data could make it possible to tailor medications to patients, thereby increasing their effectiveness and reducing their side effects. Big Data could also accelerate the research and development of new drugs and focus resources on the areas of greatest need.

Big Data applications are spreading like wildfire. They facilitate personalized offers, services and products. One of the greatest successes of Big Data is automatic speech recognition and processing. Apple's Siri understands you when asking for a restaurant, and Google Maps can lead you there. Google translate interprets foreign languages by comparing them with a huge collection of translated texts. IBM's Watson computer even understands human language. It can not only beat experienced quiz show players, but take care of customer hotlines and patients - perhaps better than humans. IBM has just decided to invest $1 billion to further develop and commercialize the system.

Of course, Big Data play an important role in the financial sector. Approximately seventy percent of all financial market transactions are now made by automated trading algorithms. In just one day, the entire money supply of the world is traded. So much money also attracts organized crime. Therefore, financial transactions are scanned by Big Data algorithms for abnormalities to detect suspicious activities. The company Blackrock uses a similar software, called

---

[3] McKinsey and Co. Open data: Unlocking innovation and performance with liquid information, http://www.mckinsey.com/insights/business_technology/open_data_unlocking_innovation_and_performance_with_liquid_information

[4] http://opensource.com/government/13/7/open-data-charter-g8
http://ec.europa.eu/digital-agenda/en/news/eu-implementation-g8-open-data-charter
http://ec.europa.eu/information_society/newsroom/cf/dae/document.cfm?doc_id=3489

"Aladdin", to successfully speculate with funds amounting approximately to the gross domestic product (GDP) of Europe.

The Big Data approach is markedly different from classical data mining approaches, where datasets have been carefully collected and carefully curated in databases by scientists or other experts. However, each year we now produce as much data as in the entire history of humankind, i.e. in all the years before. This exceeds by far human capacities to curate all data. In just one minute, 700,000 google queries and 500,000 facebook comments are sent. Besides this, enormous amounts of data are produced by all the traces that human activities are now leaving in the Internet. This includes shopping and financial data, geo-positioning and mobility data, social contacts, opinions posted in social networks, files stored in dropbox or some other cloud storage, emails posted or received through free accounts, ebooks read, including time spent on each page and sentences marked, Google or Apple Siri queries asked, youtube or TV movies watched on demand, and games played. Modern game engines and smart home equipment would also sense your activities at home, digital glasses would transmit what you see, and gene data are also massively gathered now.

Meanwhile, the data sets collected by companies such as ebay, Walmart or Facebook, reach the size of petabytes (1 million billion bytes) - one hundred times the information content of the largest library in the world: the U.S. Library of Congress. The mining of Big Data opens up entirely new possibilities for process optimization, the identification of interdependencies, and decision support. However, Big Data also comes with new challenges, which are often characterized by four criteria:

- *volume:* the file sizes and number of records are huge,
- *velocity:* the data evaluation has often to be done in real-time,
- *variety:* the data are often very heterogeneous and unstructured,
- *veracity:* the data are probably incomplete, not representative, and contain errors.

Therefore, completely new algorithms had to be developed, i.e. new computational methods.

**Machine Learning, Deep Learning, and Super-Intelligence**

To create value from data, it is crucial to turn raw data into useful information and actionable knowledge, some even aim at producing "wisdom" and "clairvoyance" (predictive capabilities). This process requires powerful computer algorithms. Machine learning algorithms do not only watch out for particular patterns, but find patterns even by themselves. This has led Chris Anderson to famously postulate "the end of theory", i.e. the hypothesis that the data deluge makes the scientific method obsolete.[5] If there would be just a big enough quantity of data, machine learning could turn it into high-quality data and come to the right conclusions. This hypothesis has become the credo of Big Data analytics, even though this almost religious belief lacks a proper foundation. I am therefore calling here for a proof of concept, by formulating the following test: Can universal machine learning algorithms, when mining huge masses of experimental physics data, discover the laws of nature themselves, without the support of human knowledge and intelligence?

---

[5] Chris Anderson, The End of Theory: The Data Deluge Makes the Scientific Method Obsolete. WIRED Magazin 16.07, http://archive.wired.com/science/discoveries/magazine/16-07/pb_theory

In spite of these issues, deep learning algorithms are celebrating great successes in everyday applications that do not require an understanding of a hidden logic or causal interdependencies.[6] These algorithms are universal learning procedures which, theoretically, could learn any pattern or input-output relation, given enough time and data. Such algorithms are particularly strong in pattern recognition tasks, i.e. reading, listening, watching, and classifying contents.[7] As a consequence, experts believe that about 50% of all current jobs in the industrial and service sectors will be lost in the next 10-20 years. Moreover, abilities comparable to the human brain are expected to be reached within the next 5 to 25 years.[8] This has led to a revival of Artificial Intelligence, now often coming under the label "Cognitive Computing".

To be competitive with intelligent machines, humans will in future increasingly need "cognitive assistants". These are digital tools such as Google Now. However, as cognitive assistants get more powerful at exponentially accelerating pace, they would soon become something like virtual colleagues, then something like digital coaches, and finally our bosses. Robots acting as bosses are already being tested.[9]

Scientists are also working on "biological upgrades" for humans. The first cyborgs, i.e. humans that have been technologically upgraded, already exist. The most well-known of them is Neil Harbisson. At the same time, there is large progress in producing robots that look and behave increasingly like humans. It must be assumed that many science fiction phantasies shown in cinemas and on TV may soon become reality.[10]

Recently, however, there are increasing concerns about artificial super-intelligences, i.e. machines that would be more intelligent than humans. In fact, computers are now better at calculating, at playing chess and most other strategic games, at driving cars, and they are performing many other specialized tasks increasingly well. Certainly, intelligent multi-purpose machines will soon exist.

---

[6] One of the leading experts in this field is Jürgen Schmidhuber.
[7] Jeremy Howard, The wonderful and terrifying implications of computers that can learn, TEDx Brussels, http://www.ted.com/talks/jeremy_howard_the_wonderful_and_terrifying_implications_of_computers_that_can_learn
[8] The point in time when this happens is sometimes called "singularity", according to Ray Kurzweil.
[9] Süddeutsche (11.3.2015) Roboter als Chef, http://www.sueddeutsche.de/leben/roboter-am-arbeitsplatz-billig-freundlich-klagt-nicht-1.2373715
[10] Such movies often serve to familiarize the public with new technologies and realities, and to give them a positive touch (including "Big Brother").

Only two or three years ago, most people would have considered it impossible that algorithms, computers, or robots would ever challenge humans as crown of creation. This has changed.[11] Intelligent machines are learning themselves, and it's now conceivable that robots build other robots that are smarter. The resulting evolutionary progress is quickly accelerating, and it is therefore just a matter of time until there are machines smarter than us. Perhaps such super-intelligences already exist. In the following, I am presenting some related quotes of some notable scientists and technology experts, who raise concerns and try to alert the public of the problems we are running into:

For example, Elon Musk of Tesla Motors voiced:[12] "I think we should be very careful about artificial intelligence. If I had to guess at what our biggest existential threat is, it's probably that. So we need to be very careful. ... I am increasingly inclined to think that there should be some regulatory oversight, maybe at the national and international level, just to make sure that we don't do something very foolish. ... "

Similar critique comes from Nick Bostrom at Oxford University.[13]

Stephen Hawking, the most famous physicist to date, recently said:[14] "Humans who are limited by slow biological evolution couldn't compete and would be superseded. ... The development of full artificial intelligence could spell the end of the human race. ... It would take off on its own, and re-design itself at an ever increasing rate."

Furthermore, Bill Gates of Microsoft was quoted:[15] "I am in the camp that is concerned about super intelligence. ... I agree with Elon Musk and some others on this and don't understand why some people are not concerned."

Steve Wozniak, co-founder of Apple, formulated his worries as follows:[16] "Computers are going to take over from humans, no question ... Like people including Stephen Hawking and Elon Musk have predicted, I agree that the future is scary and very bad for people ... If we build these devices to take care of everything for us, eventually they'll think faster than us and they'll get rid of the slow humans to run companies more efficiently ... Will we be the gods? Will we be the family pets? Or will we be ants that get stepped on? I don't know ..."

Personally, I think more positively about artificial intelligence, but I believe that we should engage in distributed collective intelligence rather than creating a few extremely powerful super-intelligences we may not be able to control.[17] It seems that various big IT companies in the Silicon Valley are already engaged in building super-intelligent machines. It was also recently reported that Baidu, the Chinese search engine, wanted to build a "China Brain Project", and was looking for significant financial contributions by the military.[18] Therefore, to be competitive, do we need to sacrifice our privacy for a society-spanning Big Data and Deep Learning project to predict the future of the world? As will become clear later on, I don't think so, because Big Data approaches and the learning of facts from the past are usually bad a predicting fundamental shifts as they occur at societal tipping points, while this is what we mainly need to care about. The combination of explanatory models with little (but right kind of) data is often superior.[19] This can deliver a better description of macro-level societal and economic change, as I will show below, and it's macro-level effects that really matter. Additionally, one should invest in tools that allow one to reveal mechanisms for the management and design of better systems. Such innovative solutions, too, cannot be found by mining data of the past and learning patterns in them.

---

[11] James Barrat (2013) Our Final Invention - Artificial Intelligence and the End of the Human Era (Thomas Dunne Books). Edge Question 2015: What do you think about machines that think? http://edge.org/annual-question/what-do-you-think-about-machines-that-think

[12] http://www.theguardian.com/technology/2014/oct/27/elon-musk-artificial-intelligence-ai-biggest-existential-threat

[13] Nick Bostrom (2014) Superintelligence: Paths, Dangers, Strategies (Oxford University Press).

[14] http://www.bbc.com/news/technology-30290540

[15] http://www.cnet.com/news/bill-gates-is-worried-about-artificial-intelligence-too/

[16] http://www.washingtonpost.com/blogs/the-switch/wp/2015/03/24/apple-co-founder-on-artificial-intelligence-the-future-is-scary-and-very-bad-for-people/

[17] D. Helbing (2015) Distributed Collective Intelligence: The Network Of Ideas, http://edge.org/response-detail/26194

**Persuasive Technologies and Nudging to Manipulate Individual Decisions**

Personal data of all kinds are now being collected by many companies, most of which are not well-known to the public. While we surf the Internet, every single click is recorded by cookies, super-cookies and other processes, mostly without our consent. These data are widely traded, even though this often violates applicable laws. By now, there are about 3,000 to 5,000 personal records of more or less every individual in the industrialized world. These data make it possible to map the way each person thinks and feels. Their clicks would not only produce a unique fingerprint identifying them (perhaps even when surfing anonymously). They would also reveal the political party they are likely to vote for (even though the anonymous vote is an important basis of democracies). Their google searches would furthermore reveal the likely actions they are going to take next (including likely financial trades[20]). There are even companies such as Recorded Future and Palantir that try to predict future individual behavior based on the data available about each of us. Such predictions seem to work pretty well, in more than 90% of all cases. It is often believed that this would eventually make the future course of our society predictable and controllable.

In the past, the attitude was "nobody is perfect, people make mistakes". Now, with the power of modern information technologies, some keen strategists hope that our society could be turned into a perfect clockwork. The feasibility of this approach is already being tested. Personalized advertisement is in fact trying to manipulate people's choices, based on the detailed knowledge of a person, including how he/she thinks, feels, and responds to certain kinds of situations.

---

[18] http://www.scmp.com/lifestyle/technology/article/1728422/head-chinas-google-wants-country-take-lead-developing, http://www.wantchinatimes.com/news-subclass-cnt.aspx?id=20150307000015&cid=1101

[19] For example, the following approach seems superior to what Google Flu Trends can offer: D. Brockmann and D. Helbing, The hidden geometry of complex, network-driven contagion phenomena. Science 342, 1337-1342 (2013).

[20] T. Preis, H.S. Moat, and H.E. Stanley, Quantifying trading behavior in financial markets using Google Trends. Scientific Reports 3: 1684 (2013).

These approaches become increasingly effective, making use of biases in human decision-making and also subliminal messages. Such techniques address people's subconsciousness, such that they would not necessarily be aware of the reasons causing their actions, similar to acting under hypnosis.

Manipulating people's choices is also increasingly being discussed as policy tool, called "nudging" or "soft paternalism".[21] Here, people's decisions and actions would be manipulated by the state through digital devices to reach certain outcomes, e.g. environmentally friendly or healthier behavior, or also certain election results. Related experiments are being carried out already.[22]

## Attempt of a Technology Assessment

In the following, I will discuss some of the social, economic, legal, ethical and other implications of the above digital technologies and their use. Like all other technologies, the use of Big Data, Artificial Intelligence, and Nudging can produce potentially harmful side effects, but in this case the impact on our economy and society may be massive. To benefit from the opportunities of digital technologies and minimize their risks, it will be necessary to combine certain technological solutions with social norms and legal regulations. In the following, I attempt to give a number of initial hints, but the discussion below can certainly not give a full account of all issues that need to be addressed.

### Problems with Big Data Analytics

The risks of Big Data are manifold. The *security* of digital communication has been undermined. Cyber crime, including data, identity and financial theft, is exploding, now producing an annual damage of the order of 3 trillion dollars, which is exponentially growing. Critical infrastructures such as energy, financial and communication systems are threatened by cyber attacks. They could, in principle, be made dysfunctional for an extended period of time, thereby seriously disrupting our economy and society. Concerns about cyber wars and digital weapons (D weapons) are quickly growing, as they may be even more dangerous than atomic, biological and chemical (ABC) weapons.

Besides cyber risks, there is a pretty long list of other problems. Results of Big Data analytics are often taken for granted and objective. This is dangerous, because the effectiveness of Big Data is sometimes based more on beliefs than on facts.[23] It is also far from clear that surveillance cameras[24] and predictive policing[25] can really significantly reduce organized and violent crime or that mass surveillance is more effective in countering terrorism than classical investigation methods.[26] Moreover, one of the key examples of the power of Big Data analytics, Google Flu Trends, has recently been found to make poor predictions. This is partly because advertisements bias user behaviors and search algorithms are being changed, such that the results are not stable and reproducible.[27] In fact, Big Data curation and calibration efforts are often low. As a consequence, the underlying datasets are typically not representative and they may contain many errors. Last but not least, Big Data algorithms are frequently used to reveal optimization potentials, but their results may be unreliable or may not reflect any causal relationships. Therefore, conclusions from Big Data are not necessarily correct.

---

[21] R.H. Thaler and C.R. Sunstein (2009) Nudge (Penguin Books).
[22] Süddeutsche (11.3.2015) Politik per Psychotrick, http://www.sueddeutsche.de/wirtschaft/verhaltensforschung-am-buerger-politik-per-psychotrick-1.2386755
[23] For example, many Big Data companies (even big ones) don't make large profits and some are even making losses. Making big money often requires to bring a Big Data company to the stock market, or to be bought by another company.
[24] M. Gill and A. Spriggs: Assessing the impact of CCTV. Home Office Research, Development and Statistics Directorate (2005), https://www.cctvusergroup.com/downloads/file/Martin%20gill.pdf; see also BBC News (August 24, 2009) 1,000 cameras `solve one crime', http://news.bbc.co.uk/2/hi/uk_news/england/london/8219022.stm

A naive application of Big Data algorithms can easily lead to mistakes and wrong conclusions. The error rate in classification problems (e.g. the distinction between "good" and "bad" risks) is often significant. Issues such as wrong decisions or discrimination are serious problems.[28] In fact, anti-discrimination laws may be implicitly undermined, as results of Big Data algorithms may imply disadvantages for women, handicapped people, or ethnic, religious, and other minorities. This is, because insurance offers, product prices of Internet shops, and bank loans increasingly depend on behavioral variables, and on specifics of the social environment, too. It might happen, for example, that the conditions of a personal loan depend on the behavior of people one has never met. In the past, some banks have even terminated loans, when neighbors have failed to make their payments on time.[29] In other words, as we lose control over our personal data, we are losing control over our lives, too. How will we then be able to take responsibility for our life in the future, if we can't control it any longer?

This brings us to the point of privacy. There are a number of important points to be considered. First of all, surveillance scares people, particularly minorities. All minorities are vulnerable, but the success of our society depends on them (e.g.

---

[25] Journalist's Resource (November 6, 2014) The effectiveness of predictive policing: Lessons from a randomized controlled trial, http://journalistsresource.org/studies/government/criminal-justice/predictive-policing-randomized-controlled-trial. ZEIT Online (29.3.2015) Predictive Policing - Noch hat niemand bewiesen, dass Data Mining der Polizei hilft, http://www.zeit.de/digital/datenschutz/2015-03/predictive-policing-software-polizei-precobs

[26] The Washington Post (January 12, 2014) NSA phone record collection does little to prevent terrorist attacks, group says, http://www.washingtonpost.com/world/national-security/nsa-phone-record-collection-does-little-to-prevent-terrorist-attacks-group-says/2014/01/12/8aa860aa-77dd-11e3-8963-b4b654bcc9b2_story.html?hpid=z4; see also http://securitydata.newamerica.net/nsa/analysis

[27] D.M. Lazer et al. The Parable of Google Flu: Traps in Big Data Analytics, Science 343, 1203-1205 (2014).

[28] D. Helbing (2015) Thinking Ahead, Chapter 10 (Springer, Berlin). See also
https://www.ftc.gov/news-events/events-calendar/2014/09/big-data-tool-inclusion-or-exclusion
https://www.whitehouse.gov/issues/technology/big-data-review
https://www.whitehouse.gov/sites/default/files/docs/Big_Data_Report_Nonembargo_v2.pdf
http://www.wsj.com/articles/SB10001424052702304178104579535970497908560

[29] This problem is related with the method of "geoscoring", see
http://www.kreditforum.net/kreditwuerdigkeit-und-geoscoring.html/

politicians, entrepreneurs, intellectuals). As the "Volkszählungsurteil"[30] correctly concludes, the continuous and uncontrolled recording of data about individual behaviors is undermining chances of personal, but also societal development. Society needs innovation to adjust to change (such as demographic, environmental, technological or climate change). However, innovation needs a cultural setting that allows to experiment and make mistakes.[31] In fact, many fundamental inventions have been made by accident or even mistake (Porcelain, for example, resulted from attempts to produce gold). A global map of innovation clearly shows that fundamental innovation mainly happens in free and democratic societies.[32] Experimenting is also needed to become an adult who is able to judge situations and take responsible decisions.

Therefore, society needs to be run in a way that is tolerant to mistakes. But today one may get a speed ticket for having been 1km/h too fast (see below). In future, in our over-regulated world, one might get tickets for almost anything.[33] Big Data would make it possible to discover and sanction any small mistake. In the USA, there are already 10 times more people in prison than in Europe (and more than in China and Russia, too). Is this our future, and does it have anything to do with the free society we used to live in? However, if we would punish only a sample of people making mistakes, how would this be compatible with fairness? Wouldn't this end in arbitrariness and undermine justice? And wouldn't the principle of assumed innocence be gone, which is based on the idea that the majority of us are good citizens, and only a few are malicious and to be found guilty?

Undermining privacy can't work well. It questions trust in the citizens, and this undermines the citizens' trust in the government, which is the basis of its legitimacy and power. The saying that "trust is good, but control is better" is not entirely correct: control cannot fully replace trust.[34] A well-functioning and efficient society needs a suitable combination of both.

"Public" without "private" wouldn't work well. Privacy provides opportunities to explore new ideas and solutions. It helps to recover from the stress of daily adaptation and reduces conflict in a dense population of people with diverse preferences and cultural backgrounds.

Public and private are two sides of the same medal. If everything is public, this will eventually undermine social norms.[35] On the long run, the consequence could be a shameless society, or if any deviation from established norms is sanctioned, a totalitarian society.

---

[30] http://de.wikipedia.org/wiki/Volksz%C3%A4hlungsurteil
[31] The Silicon Valley is well-known for this kind of culture.
[32] A. Mazloumian et al. Global multi-level analysis of the 'scientific food web', Scientific Reports 3: 1167 (2013), http://www.nature.com/srep/2013/130130/srep01167/full/srep01167.html?message-global=remove
[33] J. Schmieder (2013) Mit einem Bein im Knast - Mein Versuch, ein Jahr lang gesetzestreu zu leben (Bertelsmann).
[34] Detlef Fetchenhauer, Six reasons why you should be more trustful, TEDx Groningen, https://www.youtube.com/watch?v=gZlzCc57qX4
[35] A. Diekmann, W. Przepiorka, and H. Rauhut, Lifting the veil of ignorance: An experiment on the contagiousness of norm violations, preprint http://cess.nuff.ox.ac.uk/documents/DP2011/CESS_DP2011_004.pdf

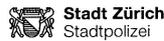

*... and I was actually not even the driver...*

Therefore, while the effects of mass surveillance and privacy intrusion are not immediately visible, they might still cause a long-term damage by undermining the fabric of our society: social norms and culture. It is highly questionable whether the economic benefits would really outweight this, and whether a control-based digital society would work at all. I rather expect such societal experiments to end in disaster.

## Problems with Artificial Intelligence and Super-Intelligence[36]

The globalization and networking of our world has caused a level of interdependency and complexity that no individual can fully grasp. This leads to the awkward situation that every one of us sees only part of the picture, which has promoted the idea that we should have artificial super-intelligences that may be able to overlook the entire knowledge of the world. However, learning such knowledge (not just the facts, but also the implications) might progress more slowly than our world changes and human knowledge progresses.[37]

---

[36] Note that super-intelligent machines may be seen as an implementation of the concept of the "wise king". However, as I am saying elsewhere, this is not a suitable approach to govern complex societies (see also the draft chapters of my book on the Digital Society at http://www.ssrn.com and https://futurict.blogspot.com, particularly the chapter on the Complexity Time Bomb: http://papers.ssrn.com/sol3/papers.cfm?abstract_id=2502559). Combinatorial complexity must be answered by combinatorial, i.e. collective intelligence, and this needs personal digital assistants and suitable information platforms for coordination.

It is also important to consider that the meaning of data depends on context. This becomes particularly clear for ambiguous content. Therefore, like our own brain, an artificial intelligence based on deep learning will sometimes see spurious correlations, and it will probably have some prejudices, too.

Unfortunately, having more information than humans (as cognitive computers have it today) doesn't mean to be objective or right. The problem of "over-fitting", according to which there is a tendency to fit meaningless, random patterns in the data is just one possible issue. The problems of parameter sensitivity or of "chaotic" or "turbulent" system dynamics will restrict possibilities to predict future events, to assess current situations, or even to identify the correct model parameters describing past events.[38] Despite these constraints, a data-driven approach would always deliver some output, but this might be just an "opinion" of an intelligent machine rather than a fact. This becomes clear if we assume to run two identical super-intelligent machines in different places. As they are not fed with exactly the same information, they would have different learning histories, and would sometimes come to different conclusions. So, super-intelligence is no guarantee to find a solution that corresponds to the truth.[39] And what if a super-intelligent machine catches a virus and gets something like a "brain disease"?

The greatest problem is that we might be tempted to apply powerful tools such as super-intelligent machines to shape our society at large. As it became obvious above, super-intelligences would make mistakes, too, but the resulting damage might be much larger and even disastrous. Besides this, super-intelligences might emancipate themselves and become uncontrollable. They might also start to act in their own interest, or lie.

Most importantly, powerful tools will always attract people striving for power, including organized criminals, extremists, and terrorists. This is particularly concerning because there is no 100% reliable protection against serious misuse. At the 2015 WEF meeting, Misha Glenny said: "There are two types of companies in the world: those that know they've been hacked, and those that don't".[40] In fact, even computer systems of many major companies, the US military, the Pentagon and the White House have been hacked in the past, not to talk about the problem of data leaks... Therefore, the growing concerns regarding building and using super-intelligences seem to be largely justified.

---

[37] Remember that it takes about 2 decades for a human to be ready for responsible, self-determined behavior. Before, however, he/she may do a lot of stupid things (and this may actually happen later, too).

[38] I. Kondor, S. Pafka, and G. Nagy, Noise sensitivity of portfolio selection under various risk measures, Journal of Banking & Finance 31(5), 1545-1573 (2007).

[39] It's quite insightful to have two phones talk to each other, using Apple's Siri assistant, see e.g. this video: https://www.youtube.com/watch?v=WuX509bXV_w

[40] http://www.brainyquote.com/quotes/quotes/m/mishaglenn564076.html, see also http://www.businessinsider.com/fbi-director-china-has-hacked-every-big-us-company-2014-10

**Problems with manipulative ("persuasive") technologies[41]**

The use of information technology is changing our behavior. This fact invites potential misuse, too.[42] Society-scale experiments with manipulative technologies are likely to have serious side effects. In particular, influencing people's decision-making undermines the principle of the "wisdom of crowds",[43] on which democratic decision-making and also the functioning of financial markets is based. For the "wisdom of crowds" to work, one requires sufficiently well educated people who gather and judge information separately and make their decisions independently. Influencing people's decisions will increase the likelihood of mistakes, which might be costly. Moreover, the information basis may get so biased over time that no one, including government institutions and intelligent machines, might be able to make reliable judgments.

Eli Pariser raised a related issue, which he called the "filter bubble". As we increasingly live in a world of personalized information, we are less and less confronted with information that doesn't fit our beliefs and taste. While this creates a feeling to live in the world we like, we will lose awareness of other people's needs and their points of view. When confronted with them, we may fail to communicate and interact constructively. For example, the US political system seems to increasingly suffer from the inability of republicans and democrats to make compromises that are good for the country. When analyzing their political discourse on certain subjects, it turns out that they don't just have different opinions, but they also use different words, such that there is little chance to develop a shared understanding of a problem.[44] Therefore, some modern information systems haven't made it easier to govern a country - on the contrary.

In perspective, manipulative technologies may be seen as attempts to "program" people. Some crowd sourcing techniques such as the services provided by Mechanical Turk come already pretty close to this. Here, people pick up jobs of all kinds, which may just take a few minutes. For example, you may have a 1000 page manual translated in a day, by breaking it down into sufficiently many micro-translation jobs.[45] In principle, however, one could think of anything, and people might not even be aware of the outcome they are jointly producing.[46]

---

[41] In other places (http://futurict.blogspot.com/2014/10/crystal-ball-and-magic-wandthe.html), I have metaphorically compared these technologies with a "magic wand" ("Zauberstab"). The problem with these technologies is: they are powerful, but if we don't use them well, their use can end in disaster. A nice poem illustrating this is The Sourcerer's Apprentice by Johann Wolfgang von Goethe: http://germanstories.vcu.edu/goethe/zauber_dual.html, http://www.rither.de/a/deutsch/goethe/der-zauberlehrling/

[42] For example, it recently became public that Facebook had run a huge experiment trying to manipulate people's mood: http://www.theatlantic.com/technology/archive/2014/09/facebooks-mood-manipulation-experiment-might-be-illegal/380717/ This created a big "shit storm": http://www.wsj.com/articles/furor-erupts-over-facebook-experiment-on-users-1404085840. However, it was also attempted to influence people's voting behavior: http://www.nzz.ch/international/kampf-um-den-glaesernen-waehler-1.18501656 OkCupid even tried to manipulate people's private emotions: http://www.theguardian.com/technology/2014/jul/29/okcupid-experiment-human-beings-dating It is also being said that each of our Web searches now triggers about 200 experiments.

[43] J. Lorenz et al. How social influence can undermine the wisdom of crowd effect, Proceedings of the National Academy of Science of the USA 108 (22), 9020-9025 (2011); see also J. Surowiecki (2005) The Wisdom of Crowds (Anchor).

[44] See Marc Smith's analyses of political discourse with NodeXL: http://nodexl.codeplex.com/

Importantly, manipulation incapacitates people, and it makes them less capable of solving problems by themselves.[47] On the one hand, this means that they increasingly lose control of their judgments and decision-making. On the other hand, who should be held responsible for mistakes that are based on manipulated decisions? The one who took the wrong decision or the one who made him or her take the wrong decision? Probably the latter, particularly as human brains can decreasingly keep up with the performance of computer systems (think, for example, of high-frequency trading).

Finally, we must be aware of another important issue. Some keen strategists believe that manipulative technologies would be perfect tools to create a society that works like a perfect machine. The idea behind this is as follows: A super-intelligent machine would try to figure out an optimal solution to a certain problem, and it would then try to implement it using punishment or manipulation, or both. In this context, one should evaluate again what purposes recent editions of security laws (such as the BÜPF) might be used for, besides fighting true terrorists. It is certainly concerning if people can be put to jail for contents on their computer hard disks, while at the same time hard disks are known to have back doors, and secret services are allowed to download materials to them. This enables serious misuse, but it also questions whether hard disk contents can be still accepted as evidence at court.

Of course, one must ask, whether it would be really possible to run a society by a combination of surveillance, manipulation and coercion. The answer is: probably yes, but given the complexity of our world, I expect this would not work well and not for long. One might therefore say that, in complex societies, the times where a "wise king" or "benevolent dictator" could succeed are gone.[48] But there is the serious danger that some ambitious people might still try to implement the concept and take drastic measures in desperate attempts to succeed. Minorities, who are often seen to produce "unnecessary complexity", would probably get under pressure.[49] This would reduce social, cultural and economic diversity.

As a consequence, this would eventually lead to a socio-economic "diversity collapse", i.e. many people would end up behaving similarly. While this may appear favorable to some people, one must recognize that diversity is the basis

---

[45] M. Bloodgood and C. Callison-Burch, Using Mechanical Turk to build machine translation evaluation sets, http://www.cis.upenn.edu/~ccb/publications/using-mechanical-turk-to-build-machine-translation-evaluation-sets.pdf

[46] In an extreme case, this might even be a criminal act.

[47] Interestingly, for IBM Watson (the intelligent cognitive computer) to work well, it must be fed with non-biased rather than with self-consistent information, i.e. pre-selecting inputs to get rid of contradictory information reduces Watson's performance.

[48] It seems, for example, that the attempts of the world's superpower to extend its powers have rather weakened it: we are now living in a multi-polar world. Coercion works increasingly less. See the draft chapters of my book on the Digital Society at http://ssrn.com for more information.

[49] even though one never knows before what kinds of ideas and social mechanisms might become important in the future - innovation always starts with minorities

of innovation, economic development,[50] societal resilience, collective intelligence, and individual happiness. Therefore, socio-economic and cultural diversity must be protected in a similar way as we have learned to protect biodiversity.[51]

Altogether, it is more appropriate to compare a social or economic system to an ecosystem than to a machine. It then becomes clear that a reduction of diversity corresponds to the loss of biological species in an ecosystem. In the worst case, the ecosystem could collapse. By analogy, the social or economic system would lose performance and become less functional. This is what typically happens in totalitarian regimes, and it often ends with wars as a result of attempts to counter the systemic instability caused by a diversity collapse.[52]

In conclusion, to cope with diversity, engaging in interoperability is largely superior to standardization attempts. That is why I am suggesting below to develop personal digital assistants that help to create benefits from diversity.

## Recommendations

I am a strong supporter of using digital technologies to create new business opportunities and to improve societal well-being. Therefore, I think one shouldn't stop the digital revolution. (Such attempts would anyway fail, given that all countries are exposed to harsh international competition.) However, like with every technology, there are also potentially serious side effects, and there is a dual use problem.

If we use digital technologies in the wrong way, it could be disastrous for our economy, ending in mass unemployment and economic depression. Irresponsible uses could also be bad for our society, potentially ending (intentionally or not) in more or less totalitarian regimes with little individual freedoms.[53] There are also serious security issues due to exponentially increasing cyber crime, which is partially related to the homogeneity of our current Internet, the lack of barriers (for the sake of efficiency), and the backdoors in many hard- and software systems.

Big Data produces further threats. It can be used to ruin personal careers and companies, but also to launch cyber wars.[54] As we don't allow anyone to own a nuclear bomb or to drive a car without breaks and other safety equipment, we must regulate and control the use of Big Data, too, including the use of Big Data by governments and secret services. This seems to require a sufficient level of transparency, otherwise it is hard to judge for anyone whether we can trust such uses and what are the dangers.

---

[50] C.A. Hidalgo et al. The product space conditions the development of nations, Science 317, 482-487 (2007). According to Jürgen Mimkes, economic progress (which goes along with an increase in complexity) also drives a transition from autocratic to democratic governance above a certain gross domestic product per capita. In China, this transition is expected to happen soon.
[51] This is the main reason why one should support pluralism.
[52] See the draft chapters of D. Helbing's book on the Digital Society at http://www.ssrn.com, particular the chapter on the Complexity Time Bomb
[53] One might distinguish these into two types: dictatorships based on surveillance ("Big Brother") and manipulatorships ("Big Manipulator").
[54] As digital weapons, so-called D-weapons, are certainly not less dangerous than atomic, biological and chemical (ABC) weapons, they would require international regulation and control.

**Recommendations regarding Big Data**

The use of Big Data should meet certain quality standards. This includes the following aspects:
- Security issues must be paid more attention to. Sensitive data must be better protected from illegitimate access and use, including hacking of personal data. For this, more and better data encryption might be necessary.
- Storing large amounts of sensitive data in one place, accessible with a single password appears to be dangerous. Concepts such as distributed data storage and processing are advised.
- It should not be possible for a person owning or working in a Big Data company or secret service to look into personal data in unauthorized ways (think of the LoveINT affair, where secret service staff was spying on their partners or ex-partners[55]).
- Informational self-determination (i.e. the control of who uses what personal data for what purpose) is necessary for individuals to keep control of their lives and be able to take responsibility for their actions.
- It should be easy for users to exercise their right of informational self-determination, which can be done by means of Personal Data Stores, as developed by the MIT[56] and various companies. Microsoft seems to be working on a hardware-based solution.
- It must be possible and reasonably easy to correct wrong personal data.
- As Big Data analytics often results in meaningless patterns and spurious correlations, for the sake of objectivity and in order to come to reliable conclusions, it would be good to view its results as hypotheses and to verify or falsify them with different approaches afterwards.
- It must be ensured that scientific standards are applied to the use of Big Data. For example, one should require the same level of significance that is demanded in statistics and for the approval of medical drugs.
- The reproducibility of results of Big Data analytics must be demanded.
- A sufficient level of transparency and/or independent quality control is needed to ensure that quality standards are met.
- It must be guaranteed that applicable antidiscrimination laws are not implicitly undermined and violated.
- It must be possible to challenge and check the results of Big Data analytics.
- Efficient procedures are needed to compensate individuals and companies for improper data use, particularly for unjustified disadvantages.

---

[55] see http://www.washingtonpost.com/blogs/the-switch/wp/2013/08/24/loveint-when-nsa-officers-use-their-spying-power-on-love-interests/
[56] http://openpds.media.mit.edu/

- Serious violations of constitutional rights and applicable laws should be confronted with effective sanctions.
- To monitor potential misuse and for the sake of transparency, the processing of sensitive data (such as personal data) should probably be always logged.
- Reaping private benefits at the cost of others or the public must be sanctioned.
- As for handling dangerous goods, potentially sensitive data operations should require particular qualifications and a track record of responsible behavior (which might be implemented by means of special kinds of reputation systems).

It must be clear that digital technologies will only thrive if they are used in a responsible way. For companies, the trust of consumers and users is important to gain and maintain a large customer base. For governments, public trust is the basis of legitimacy and power. Losing trust would, therefore, cause irrevocable damage.

The current problem is the use of cheap technology. For example, most software is not well tested and not secure. Therefore, Europe should invest in high-quality services of products. Considering the delay in developing data products and services, Europe must anyway find a strategy that differentiates itself from its competitors (which could include an open data and open innovation strategy, too[57]).

Most, if not all functionality currently produced with digital technologies (including certain predictive "Crystal Ball" functionality[58]) can be also obtained in different ways, particularly in ways that are compatible with constitutional and data protection laws[59] (see also the Summary, Conclusion, and Discussion). This may come at higher costs and slightly reduced efficiency, but it might be cheaper overall than risking considerable damage (think of the loss of 3 trillion dollars by cybercrime each year and consider that this number is still exponentially increasing). Remember also that we have imposed safety requirements on nuclear, chemical, genetic, and other technologies (such as cars and planes) for good reasons. In particular, I believe that we shouldn't (and wouldn't need to) give up the very important principle of informational self-determination in order to unleash the value of personal data. Informational self-control is of key importance to keep democracy, individual freedom, and responsibility for our lives. To reach catalytic and synergy effects, I strongly

---

[57] http://horizon-magazine.eu/article/open-data-could-turn-europe-s-digital-desert-digital-rainforest-prof-dirk-helbing_en.html, https://ec.europa.eu/digital-agenda/en/growth-jobs/open-innovation

[58] http://www.defenseone.com/technology/2015/04/can-military-make-prediction-machine/109561/

[59] D. Helbing and S. Balietti, From social data mining to forecasting socio-economic crises, Eur. Phys. J Special Topics 195, 3-68 (2011); see also http://www.zeit.de/digital/datenschutz/2015-03/datenschutzverordnung-zweckbindung-datensparsamkeit; http://www.google.com/patents/US8909546

advise to engage in culturally fitting uses of Information and Communication Technologies (ICT).

In order to avoid slowing down beneficial data applications too much, one might think of continuously increasing standards. Some new laws and regulations might become applicable within 2 or 3 years time, to give companies a sufficiently long time to adjust their products and operation. Moreover, it would be useful to have some open technology standards such that all companies (also small and medium-sized ones) have a chance to meet new requirements with reasonable effort. Requiring a differentiated kind of interoperability could be of great benefit.

**Recommendations regarding Machine Learning and Artificial Intelligence**

Modern data application go beyond Big Data analytics towards (semi-)automatic systems, which typically offer possibilities for users to control certain system parameters (but sometimes there is just the possibility to turn the automatic or the system off). Autopilot systems, high-frequency trading, and self-driving cars are well-known examples. Would we in future even see an automation of society, including an automated voting by digital agents mirroring ourselves?[60]

Automated or autonomous systems are often not a 100 percent controllable, as they may operate at a speed that humans cannot compete with. One must also realize that today's artificial intelligent systems are not fully programmed. They learn, and they may therefore behave in ways that have not been tested before. Even if their components would be programmed line by line and would be thoroughly tested without showing any signs of error, the interaction of the system components may lead to unexpected behaviors. For example, this is often the case when a car with sophisticated electronic systems shows surprising behavior (such as suddenly not operating anymore). In fact, unexpected ("emergent") behavior is a typical feature of many complex dynamical systems.

The benefits of intelligent learning systems can certainly be huge. However, we must understand that they will sometimes make mistakes, too, even when automated systems are superior to human task performance. Therefore, one should make a reasonable effort to ensure that mistakes by an automated system are outweighted by its benefits. Moreover, possible damages should be sufficiently small or rare, i.e. acceptable to society. In particular, such damages should not pose any large-scale threats to critical infrastructures, our economy, or our society. As a consequence, I propose the following:

- A legal framework for automated technologies and intelligent machines is necessary. Autonomy needs to come with responsibility, otherwise one may quickly end in anarchy and chaos.

---

[60] https://www.youtube.com/watch?v=mO-3yVKuDXs , https://www.youtube.com/watch?v=KgVBob5HIm8

- Companies should be accountable for delivering automated technologies that satisfy certain minimum standards of controllability and for sufficiently educating their users (if necessary).
- The users of automated technologies should be accountable for appropriate efforts to control and use them properly.
- Contingency plans should be available for the case where an automated system gets out of control. It would be good to have a fallback level or plan B that can maintain the functionality of the system at the minimum required performance level.
- Insurances and other legal or public mechanisms should be put in place to appropriately and efficiently compensate those who have suffered damage.
- Super-intelligences must be well monitored and should have in-built destruction mechanisms in case they get out of control nevertheless.
- Relevant conclusions of super-intelligent systems should be independently checked (as these could also make mistakes, lie, or act selfishly). This requires suitable verification methods, for example, based on collective intelligence. Humans should still have possibilities to judge recommendations of super-intelligent machines, and to put their suggestions in a historical, cultural, social, economic and ethical perspective.
- Super-intelligent machines should be accessible not only to governing political parties, but also to the opposition (and their respectively commissioned experts), because the discussion about the choice of the goal function and the implication of this choice is inevitable. This is where politics still enters in times of evidence- or science-based decision-making.
- The application of automation should affect sufficiently small parts of the entire system only, which calls for decentralized, distributed, modular approaches and engineered breaking points to avoid cascade effects. This has important implications for the design and management of automated systems, particularly of globally coupled and interdependent systems.[61]
- In order to stay in control, governments must regulate and supervise the use of super-intelligences with the support of qualified experts and independent scientists.[62]

**Recommendations regarding manipulative technologies**

Manipulative technologies are probably the most dangerous among the various digital technologies discussed in this paper, because we might not even notice the manipulation attempts.

In the past, we lived in an information-poor world. Then, we had enough time to assess the value of information, but we did not always have enough information to decide well. With more information (Web search, Wikipedia, digital maps, etc.)

---

[61] Note that the scientific field of complexity science has a large fundus of knowledge how to reach globally coordinated results based on local interactions.
[62] After all, humans have to register, too.

orientation is increasingly easy. Now, however, we are faced with a data deluge and are confronted with so much information that we can't assess and process it all. We are blinded by too much information, and this makes us vulnerable to manipulation. We increasingly need information filters, and the question is, who should produce these information filters? A company? Or the state?

In both cases, this might have serious implications for our society, because the filters would pursue particular interests (e.g. to maximize clicks on ads or to manipulate people in favor of nationalism). In this way, we might get stuck in a "filter bubble".[63] Even if this filter bubble would feel like a golden cage, it would limit our imagination and capacity of innovation. Moreover, mistakes can and will always happen, even if best efforts to reach an optimum outcome are made.

While some problems can be solved well in a centralized fashion (i.e. in a top-down way), some optimization problems are notoriously hard and better solved in a distributed way. Innovation is one of these areas.[64] The main problem is that *the most fundamental question of optimization is unsolved, namely what goal function to choose.* When a bad goal function is chosen, this will have bad outcomes, but we may notice this only after many years. As mistakes in choosing the goal function will surely sometimes happen, it could end in disaster when everyone applies the same goal function.[65]

Therefore, one should apply something like a portfolio strategy. Under strongly variable and hardly predictable conditions a diverse strategy works best. Therefore, pluralistic information filtering is needed. In other words, customers, users, and citizens should be able to create, select, share and adapt the information filters they use, thereby creating an evolving ecosystem of increasingly better filters. In fact, everyone would probably be using several different filters (for example, "What's currently most popular?", "What's most controversial?", "What's trendy in my peer group?", "Surprise me!").[66] In contrast, if we leave it to a company or the state to decide how we see the world, we might happen to end up with biased views, and this could lead to terrible mistakes. This could, for example, undermine the "wisdom of crowds", which is currently the basis of free markets and democracies (with benefits such as a high level of performance [not necessarily growth], quality of life, and the avoidance of mistakes such as wars among each other).

In a world characterized by information overload, unbiased and reliable information becomes ever more important. Otherwise the number of mistakes will probably increase. For the digital society to succeed, we must therefore take safeguards against information pollution and biases. Reputation systems might be a suitable instrument, if enough information providers compete efficiently with each other for providing more reliable and more useful information. Additionally, legal sanctions might be necessary to counter intentionally misleading information.

---

[63] E. Pariser (2012) The Filter Bubble: How the New Personalized Web Is Changing What We Read and How We Think (Penguin).
[64] Some problems are so hard that no government and no company in the world have solved them (e.g. how to counter climate change). Large multi-national companies are often surprisingly weak in delivering fundamental innovations (probably because they are too controlling). That's why they keep buying small and medium-sized companies to compensate for this problem.
[65] Similar problems are known for software products that are used by billions of people: a single software bug can cause large-scale problems - and the worrying vulnerability to cyber attacks is further increasing.
[66] We have demonstrated such an approach in the Virtual Journal platform (http://vijo.inn.ac)

Consequently, advertisements should be marked as such, and the same applies to manipulation attempts such as nudging. In other words, the user, customer or citizen must be given the possibility to consciously decide for or against a certain decision or action, otherwise individual autonomy and responsibility are undermined. Similarly as customers of medical drugs are warned of potential side effects, one should state something like "This product is manipulating your decisions and is trying to make you behave in a more healthy way (or in a environmentally friendly way, or whatever it tries to achieve...)". The customer would then be aware of the likely effects of the information service and could actively decide whether he or she wants this or not.

Note, however, that it is currently not clear what the side effects of incentivizing the use of manipulative technologies would be. If applied on a large scale, it might be almost as bad as hidden manipulation. Dangerous herding effects might occur (including mass psychology as it occurs in hypes, stock market bubbles, unhealthy levels of nationalism, or the particularly extreme form it took during the Third Reich). Therefore,

- manipulation attempts should be easily recognizable, e.g. by requiring everyone to mark the kind of information (advertisement, opinion, or fact),
- it might be useful to monitor manipulation attempts and their effects,
- the effect size of manipulation attempts should be limited to avoid societal disruptions,
- one should have a possibility to opt out for free from the exposure to manipulative influences,
- measures to ensure pluralism and socio-economic diversity should be required,
- sufficiently many independent information providers with different goals and approaches would be needed to ensure an effective competition for more reliable information services,
- for collective intelligence to work, having a knowledge base of trustable and unbiased facts is key, such that measures against information pollution are advised.[67]

Ethical guidelines, demanding certain quality standards, and sufficient transparency might also be necessary. Otherwise, the large-scale application of manipulative technologies could intentionally or unintentionally undermine the individual freedom of decision-making and the basis of democracies, particularly

---

[67] In fact, to avoid mistakes, the more we are flooded with information the more must we be able to rely on it, as we have increasingly less time to judge its quality.

when nudging techniques become highly effective and are used to manipulate public opinion at large.[68]

## Summary, Conclusions and Discussion

Digital technologies offer great benefits, but also substantial risks. They may help us to solve some long-standing problems, but they may also create new and even bigger issues. In particular, if wrongly used, individual autonomy and freedom, responsible decision-making, democracy and the basis of our legal system are at stake. The foundations on which our society is build might be damaged intentionally or unintentionally within a very short time period, which may not give us enough opportunities to prepare for or respond to the challenges.

Currently, some or even most Big Data practices violate applicable data protection laws. Of course, laws can be changed, but some uses of Big Data are also highly dangerous, and incompatible with our constitution and culture. These challenges must be addressed by a combination of technological solutions (such as personal data stores), legal regulations, and social norms. Distributed data, distributed systems and distributed control, sufficiently many competitors and suitably designed reputation systems might be most efficient to avoid misuses of digital technologies, but transparency must be increased as well.

Even though our economy and society will change in the wake of the digital revolution, we must find a way that is consistent with our values, culture, and traditions, because this will create the largest synergy effects. In other words, a China or Singapore model is unlikely to work well in Europe.[69] We must take the next step in our cultural, economic and societal evolution.

I am convinced that it is now possible to use digital technologies in ways that bring the perspectives of science, politics, business, society, cultural traditions, ethics, and perhaps even religion together.[70] Specifically, I propose to use the Internet of Things as basis for a participatory information system called the Planetary Nervous System or Nervousnet, to support tailored measurements, awareness, coordination, collective intelligence, and informational self-determination.[71] The system I suggest would have a resilient systems design and could be imagined as a huge catalyst of socio-economic value generation. It would also support real-time feedbacks through a multi-dimensional exchange system ("multi-dimensional finance"). This approach would allow one to massively increase the efficiency of many systems, as it would support the self-organization of structures, properties and functions that we would like to have, based on local interactions. The distributed approach I propose is consistent with individual autonomy, free decision-making, the democratic principle of participation, as well as free entrepreneurial activities and markets. In fact, wealth is not only created by producing economies of scale (i.e. cheap mass production), but also by engaging in social interaction (that's why cities are drivers of the economy[72]).

---

[68] This could end up in a way of organizing our society that one could characterize as "Big Manipulator" (to be distinguished from "Big Brother").

[69] The following recent newspaper articles support this conclusion:
http://www.zeit.de/politik/ausland/2015-03/china-wachstum-fuenf-vor-acht ,
http://bazonline.ch/wirtschaft/konjunktur/China-uebernimmt-die-rote-Laterne/story/20869017 , http://www.nzz.ch/international/asien-und-pazifik/singapurer-zeitrechnung-ohne-lee-kuan-yew-1.18510938. In fact, based on a statistical analysis of Jürgen Mimkes and own observations, I expect that China will now undergo a major transformation towards a more democratic state in the coming years. First signs of instability of the current autocratic system are visible already, such as the increased attempts to control information flows.

[70] D. Helbing, Responding to complexity in socio-economic systems: How to build a smart and resilient society? Preprint http://papers.ssrn.com/sol3/papers.cfm?abstract_id=2583391

[71] D. Helbing, Creating ("Making") a Planetary Nervous System as Citizen Web, http://futurict.blogspot.jp/2014/09/creating-making-planetary-nervous.html

The proposed approach would also consider (and potentially trade) externalities, thereby supporting other-regarding and fair solutions, which would be good for our environment, too. Finally, everyone could reap the benefits of diversity by using personal digital assistants, which would support coordination and cooperation of diverse actors and reducing conflict.

In conclusion, we have the choice between two kinds of a digital society: (1) a society in which people are expected to obey and perform tasks like a robot or a gearwheel of a perfect machine, characterized by top-down control, limitations of freedom and democracy, and potentially large unemployment rates; (2) a participatory society with space for humans with sometimes surprising behaviors characterized by autonomous but responsible decision-making supported by personal digital assistants, where information is opened up to everyone's benefits in order to reap the benefits of diversity, creativity, and exponential innovation. What society would you choose?

The FuturICT community (www.futurict.eu) has recently worked out a framework for a smart digital society, which is oriented at international leadership, economic prosperity, social well-being, and societal resilience, based on the well-established principle of subsidiarity. With its largely distributed, decentralized approach, it is designed to cope with the complexity of our globalized world and benefit from it.[73]

The FuturICT approach takes the following insights into account:
- Having and using more data is not always better (e.g. due to the problem of "over-fitting", which makes conclusions less useful).[74]
- Information always depends on context (and missing context), and it is therefore never objective. One person's signal may be another person's noise and vice versa. It all depends on the question and perspective.[75]

---

- Even if individual decisions can be correctly predicted in 96% of all cases, this does not mean that the macro-level outcome would be correctly predicted.[76] This surprising discovery applies to cases of unstable system dynamics, where minor variations can lead to completely different outcomes.[77]
- In complex dynamical systems with many interacting components, even the perfect knowledge of all individual component properties does not necessarily allow one to predict what happens if components interact.[78]
- What governments really need to pay attention to are macro-effects, not micro-behavior. However, the macro-dynamics can often be understood by means of models that are based on aggregate variables and parameters.
- What matters most is whether a system is stable or unstable. In case of stability, variations in individual behavior do not make a significant difference, i.e. we don't need to know what the individuals do. In case of instability, random details matter, such that the predictability is low, and even in the unlikely case that one can exactly predict the course of events, one may not be able to control it because of cascade-effects in the system that exceed the control capacities.[79]
- Surprises and mistakes will always happen. This can disrupt systems, but many inventions wouldn't exist, if this wasn't the case.[80]
- Our economy and society should be organized in a way that manages to keep disruptions small and to respond flexibly to surprises of all kinds. Socio-economic systems should be able to resist shocks and recover from them quickly and well. This is best ensured by a resilient system design.[81]

---

[76] M. Maes and D. Helbing, Noise can improve social macro-predictions when micro-theories fail, preprint.

[77] We know this also from so-called "phantom traffic jams", which appear with no reason, when the car density exceeds a certain critical value beyond which traffic flow becomes unstable. Such phantom traffic jams could not be predicted at all by knowing all drivers thoughts and feelings in detail. However, they can be understood for example with macro-level models that do not require micro-level knowledge. These models also show how traffic congestion can be avoided: by using driver assistance systems that change the interactions between cars, using real-time information about local traffic conditions. Note that this is a distributed control strategy.

[78] Assume one knows the psychology of two persons, but then they accidentally meet and fall in love with each other. This incident will change their entire lives, and in some cases it will change history too (think of Julius Caesar and Cleopatra, for example, but there are many similar cases). A similar problem is known from car electronics: even if all electronic components have been well tested, their interaction often produces unexpected outcomes. In complex systems, such unexpected, "emergent" system properties are quite common.

[79] In case of cascade effects, a local problem will cause other problems before the system recovers from the initial disruption. Those problems trigger further ones, etc. Even hundreds of policemen could not avoid phantom traffic jams from happening, and in the past even large numbers of security forces have often failed to prevent crowd disasters (they have sometimes even triggered or deteriorated them while trying to avoid them), see D. Helbing and P. Mukerji, Crowd disasters as systemic failures: Analysis of the Love Parade disaster, EPJ Data Science 1:7 (2012).

[80] I am personally convinced that the level of randomness and unpredictability in a society is relatively high, because it creates a lot of personal and societal benefits, such as creativity and innovation. Also think of the success principle of serendipity.

[81] D. Helbing et al. FuturICT: Participatory computing to understand and manage our complex world in a more sustainable and resilient way. Eur. Phys. J. Special Topics 214, 11-39 (2012).

- A more intelligent machine is not necessarily more useful. Distributed collective intelligence can better respond to the combinatorial complexity of our world.[82]
- In complex dynamical systems which vary a lot, are hard to predict and cannot be optimized in real-time (as it applies to NP-hard control problems such as traffic light optimization), distributed control can outperform top-down control attempts by flexibly adapting to local conditions and needs.
- While distributed control may be emulated by centralized control, a centralized approach might fail to identify the variables that matter.[83] Depending on the problem, centralized control is also considerably more expensive, and it tends to be less efficient and effective.[84]
- Filtering out information that matters is a great challenge. Explanatory models that are combined with little, but the right kind of data are best to inform decision-makers. Such models also indicate what kind of data is needed.[85] Finding the right models typically requires interdisciplinary collaborations, knowledge about complex systems, and open scientific discussions that take all relevant perspectives on board.
- Diversity and complexity are not our problem. They come along with the socio-economic and cultural evolution. However, we have to learn how to use complexity and diversity to our advantage. This requires the understanding of the hidden forces behind socio-economic change, the use of (guided) self-organization and digital assistants to create interoperability and to support the coordination of actors with diverse interests and goals.
- To catalyze the best outcomes and create synergy effects, information systems should be used in a culturally fitting way.[86]
- Responsible innovation, trustable systems and a sufficient level of transparency and democratic control can be highly beneficial.

As a consequence of the above insights, to reap the benefits of data, I believe we do not need to end privacy and informational self-determination. The best use of information systems is made, if they boost our society and economy to full capacity, i.e. if they use the knowledge, skills, and resources of everyone in the best possible way. This is of strategic importance and requires suitably designed participatory information systems, which optimally exploit the special properties of information.[87] In fact, the value of participatory systems, as pointed out by Jeremy Rifkin[88] and others,[89] becomes particularly clear if we think of the great success of crowd sourcing (Wikipedia, OpenStreetMap, Github, etc.), crowd funding, citizen science and collective ("swarm") intelligence. So, let's build these systems together. What are we waiting for?

---

[82] As we know, intellectual discourse can be a very effective way of producing new insights and knowledge.

[83] Due to the data deluge, the existing amounts of data increasingly exceed the processing capacities, which creates a "flashlight effect": while we might look at anything, we need to decide what data to look at, and other data will be ignored. As a consequence, we often overlook things that matter. While the world was busy fighting terrorism in the aftermath of September 11, it did not see the financial crisis coming. While it was focused on this, it did not see the Arab Spring coming. The crisis in Ukraine came also as a surprise, and the response to Ebola came half a year late. Of course, the possibility or likelihood of all these events was reflected by some existing data, but we failed to pay attention to them.

[84] The classical telematics solutions based on a control center approach haven't improved traffic much. Today's solutions to improve traffic flows are mainly based on distributed control approaches: self-driving cars, intervehicle communication, car-to-infrastructure communication etc.

[85] This approach corresponds exactly how Big Data are used at the elementary particle accelerator CERN; 99.9% of measured data are deleted immediately. One only keeps data that are required to answer a certain question, e.g. to validate or falsify implications of a certain theory.

[86] J. van den Hoven et al. FuturICT - The road towards ethical ICT, Eur. Phys. J. Special Topics 214, 153-181 (2012).

---

[87] This probably requires different levels of access depending on qualification, reputation, and merit.
[88] J. Rifkin (2013) The Third Industrial Revolution (Palgrave Macmillan Trade); J. Rifkin (2014) The Zero Marginal Cost Society (Palgrave Macmillan Trade).
[89] Government 3.0 initiative of the South Korean government, http://www.negst.com.ng/documents/Governing_through_Networks/3-icegov2013_submission_19.pdf
http://www.koreaittimes.com/story/32400/government-30-future-opening-sharing-communication-and-collaboration